\documentclass[preprint2]{aastex}

\shorttitle{PN progenitors in dwarf irregulars}
\shortauthors{Richer \& McCall}

\begin{document}


\title{The Progenitors of Planetary Nebulae in Dwarf Irregular Galaxies}


\author{Michael G. Richer\footnote{Visiting Astronomer, Canada-France-Hawaii Telescope, operated by the National Research Council of Canada, le Centre National de la Recherche Scientifique de France, and the University of Hawaii}}
\affil{OAN, Instituto de Astronom\'\i a, Universidad Nacional Aut\'onoma de M\'exico, P.O. Box 439027, San Diego, CA 92143}
\email{richer@astrosen.unam.mx}

\and

\author{Marshall L. McCall$^1$}
\affil{Department of Physics \& Astronomy, York University, 4700 Keele Street, Toronto, Ontario, Canada  M3J 1P3}
\email{mccall@yorku.ca}



\begin{abstract}

We present chemical abundances for planetary nebulae and \ion{H}{2} regions in the Local Group dwarf irregular galaxy NGC 6822 based upon spectroscopy obtained at the Canada-France-Hawaii Telescope using the Multi-Object Spectrograph.  From these and similar data compiled from the literature for planetary nebulae in the Magellanic Clouds, Sextans A, Sextans B, and Leo A, we consider the origin and evolution of the stellar progenitors of bright planetary nebulae in dwarf irregular galaxies.  On average, the oxygen abundance observed in the bright planetary nebulae in these galaxies coincides with that measured in the interstellar medium, indicating that, in general, the bright planetary nebulae in dwarf irregulars descend primarily, though not exclusively, from stars formed in the relatively recent past.  We also find that the ratio of neon to oxygen abundances in these bright planetary nebulae is identical to that measured in the interstellar medium, indicating that neither abundance is significantly altered as a result of the evolution of their stellar progenitors.  We do find two planetary nebulae, that in Sextans A and S33 in NGC 6822, where oxygen appears to have been dredged up, but these are the exception rather than the rule.  In fact, we find that even nitrogen is not always dredged up, so it appears that the dredge-up of oxygen is uncommon for the abundance range of the sample.

\end{abstract}


%% subject header.  Objects should be in the appropriate "individual"
%% headers (e.g. quasars: individual, stars: individual, etc.) with the
%% additional provision that the total number of headers, including each
%% individual object, not exceed six.  The \objectname{} macro, and its
%% alias \object{}, is used to mark each object.  The macro takes the object
%% name as its primary argument.  This name will appear in the paper
%% and serve as the link's anchor in the electronic edition if the name
%% is recognized by the data centers.  The macro also takes an optional
%% argument in parentheses in cases where the data center identification
%% differs from what is to be printed in the paper.



\section{Introduction}

It is well-known that the stellar progenitors of planetary nebulae modify their initial chemical composition through the course of their evolution.  Typically, the matter that is returned to the interstellar medium via the nebular shell is enriched in helium and often in nitrogen, carbon, and s-process elements \citep[e.g., ][]{forestinicharbonnel1997, marigo2001}.  Recently, theoretical work has suggested that oxygen is also produced as a result of dredge-up in the progenitors of planetary nebulae from very low metallicities to those as high as that of the LMC \citep[e.g., ][]{marigo2001,herwig2004}.  These models indicate that the dredge-up of oxygen should be a relatively common result of the third dredge-up that accompanies thermal pulses on the asymptotic giant branch, at least at low metallicities (e.g., $Z < 0.25 Z_{\sun}$), especially for planetary nebulae derived from progenitors in the $2-3\,M_{\sun}$ range.  However, clear examples of its occurrence are rare.   \citet{pequignotetal2000} found evidence of oxygen dredge-up in the planetary nebula He 2-436 in the Sagittarius dwarf spheroidal from a detailed comparison with Wray 16-423, another planetary nebula in the same galaxy.  

In principle, planetary nebulae in dwarf irregulars allow the hypothesis that  oxygen is dredged up to be tested rather carefully, since the limiting composition of the stellar progenitors is known, namely that of the interstellar medium.  At least in the Magellanic Clouds, Sextans B, and Leo A, bright planetary nebulae have oxygen abundances similar to those found in the interstellar medium \citep{richer1993, magrinietal2005, kniazevetal2005, vanzeeetal2006}.  Preliminary results indicate that this situation also holds in NGC 6822 and NGC 3109 \citep{kniazevetal2006, leisyetal2006}.  The evidence currently available also indicates that the progenitors of these planetary nebulae generally did not modify their initial oxygen abundances significantly \citep[\S 3; ][]{richer2006}.  Consequently, the high oxygen abundance in bright planetary nebulae in dwarf irregulars implies that their stellar progenitors are derived from relatively recent star formation.  If so, some of the progenitors of these planetary nebulae may have been relatively massive, in which case the chances of observing the dredge-up of nitrogen and, perhaps, oxygen should be high.  

Here, we present an analysis of nitrogen and oxygen production by considering the chemical compositions of the bright planetary nebulae in the Magellanic Clouds, Sextans A and B, NGC 6822, and Leo A.  In Section 2, we present our observations of planetary nebulae and \ion{H}{2} regions in NGC 6822 as well as the reduction and analysis of these data.   In Section 3, we compile similar data for the planetary nebulae in the Magellanic Clouds, Sextans A and B, and Leo A, and demonstrate that the abundances in planetary nebulae and \ion{H}{2} regions in all of these galaxies are generally similar.  In Section 4, we consider the nucleosynthetic production in the progenitors of bright planetary nebulae, finding that oxygen production is rare and that even nitrogen production is not as common as supposed.  We present our conclusions in Section 5.

\section{Observations and Analysis}

\subsection{Observations and Data Reductions}

Our observations of NGC 6822 were obtained at the Canada-France-Hawaii Telescope using the Multi-Object Spectrograph \citep[MOS;][]{lefevreetal1994} on UT 17-19 August 1998.  The MOS is a multi-object, imaging, grism spectrograph that uses focal plane masks constructed from previously acquired images.  The object slits were chosen to be 1\arcsec\ wide, but of varying lengths to accommodate the objects of interest.  In all cases, however, the slit lengths exceeded 12\arcsec.  We used a 600\,l\,mm$^{-1}$ grism that gave a dispersion of 105\,\AA\,mm$^{-1}$ and a central wavelength of 4950\,\AA.  The detector  was the STIS2 $2048 \times 2048$ CCD.  The pixel size was 21\,microns, yielding a spatial scale at the detector of 0\farcs 43\,pix$^{-1}$.  The spectral coverage varied depending upon the object's position within the field of view, but usually covered the 3700-6700\,\AA\ interval.  Likewise, the dispersion was typically 2.1\,\AA\,pix$^{-1}$, but varied depending upon the object's position within the field of view (from 1.9\,\AA\,pix$^{-1}$ to 2.2\,\AA\,pix$^{-1}$).  Images in the light of [\ion{O}{3}]$\lambda$5007 and the continuum at 5500\AA\ were acquired on UT 17 August 1998 and used to select objects for observation.  Spectra were obtained the following two nights.  A total of five exposures of 1800 seconds each were obtained.  The photometric calibration was achieved using observations of BD+17$^\circ$4708, BD+25$^\circ$3941, BD+26$^\circ$2606, BD+33$^\circ$2642, and HD 19445.  Spectra of the HgNeAr lamps were used to calibrate in wavelength.  Pixel-to-pixel variations were removed using spectra of the halogen lamp taken through the masks created for the standard stars and NGC 6822.

The spectra were reduced using the specred package of the Image Reduction and Analysis Facility\footnote{IRAF is distributed by the National Optical Astronomical Observatories, which is operated by the Associated Universities for Research in Astronomy, Inc., under contract to the National Science Foundation.} (IRAF).  The procedure for data reduction followed approximately that outlined in \citet{masseyetal1992}.  The mean of the overscan section was subtracted from all images.  A zero-correction image was constructed from overscan-subtracted bias images and was subtracted from all images.  The fit1d task was used to fit the columns of the overscan- and zero-correction-subtracted flat field images with a many-piece linear spline to remove the shape of the halogen lamp to form normalized flat field images.  Images of the standard stars and NGC 6822 were divided by the appropriate normalized flat field image.  The spectra of the planetary nebulae and \ion{H}{2} regions in NGC 6822 were then extracted and calibrated in wavelength using the spectra of the arc lamps.  Next, the spectra of the standard stars were used to calibrate in flux.  Finally, the flux- and wavelength-calibrated spectra were averaged to produce the final spectra.  

\subsection{Line Intensities}

Table \ref{table_line_int} presents the raw line intensities normalized to H$\beta$ and their uncertainties ($1\,\sigma$) for each object, measured on a scale where the H$\beta$ line has an intensity of 100.  The line intensities were measured using the software described in \citet{mccalletal1985}.  This software simultaneously fits a sampled Gaussian profile to the emission line(s) and a straight line to the continuum.  The quoted errors include the uncertainties from the fit to the line itself, the fit to the reference line (H$\beta$), and the noise in the continuum for both lines.  For those lines where no uncertainty is quoted, the intensity given is an upper limit ($2\,\sigma$).  

Table \ref{table_line_int} also includes the reddening determined from $\mathrm H\alpha/\mathrm H\beta$ and $\mathrm H\gamma/\mathrm H\beta$, assuming intrinsic ratios appropriate for the temperature and density observed.  The temperature and density finally adopted for each object is given in Table \ref{table_chem_abun} (see \S 2.4 for details).  Normally, both estimates of the reddening agree within errors.  The \citet{fitzpatrick1999} reddening law was used, parametrized with a total-to-selective extinction of 3.041.  This parametrization delivers a true ratio of total-to-selective extinction of 3.07 when integrated over the spectrum of Vega \citep{mccall2004}, which is the average value for the diffuse component of the interstellar medium of the Milky Way \citep{mccallarmour2000}.  The uncertainties quoted for reddenings, temperatures, and densities are derived from the maximum and minimum line ratios allowed considering the uncertainties in the line intensities.  

The line intensities were corrected for reddening according to 
\begin{equation}
\log \frac{F(\lambda)}{F(\mathrm H\beta)} = \log \frac{I(\lambda)}{I(\mathrm H\beta)} -0.4 E(B-V)(A_1(\lambda)-A_1(\mathrm H\beta))
\end{equation}

\noindent where $F(\lambda)/F(\mathrm H\beta)$ and $I(\lambda)/I(\mathrm H\beta)$ are the observed and reddening-corrected line intensity ratios, respectively, $E(B-V)$ is the reddening determined from the $F(\mathrm H\alpha)/F(\mathrm H\beta)$ ratio, when available, and $A_1(\lambda)$ is the extinction in magnitudes for $E(B-V)=1$\,mag, i.e., $A_1(\lambda) = A(\lambda)/E(B-V)$, $A(\lambda)$ being the reddening law \citep{fitzpatrick1999}.  The values of $A_1(\lambda)$ used for all lines are given in column 3 of Table \ref{table_line_int}.  The optical depth at $1 \, \rm \mu m$, which is a much clearer descriptor of the amount of obscuration, can be computed by multiplying $E(B-V)$ by 1.054 \citep[see][]{mccall2004}.

\subsection{Object Identification and Classification}

Object names are taken from \citet{hodge1977}, \citet{killendufour1982}, \citet{hodgeetal1988}, and \citet{leisyetal2005a}.  Only the object denoted pn-020 is not included among the sources found in previous studies.  This new object is superposed on the \ion{H}{2} region Ho15.  Below, we argue that this object is a planetary nebula, so we denote it as pn-020, extending the naming scheme of \citet{leisyetal2005a}.

We classified the objects as planetary nebulae if \ion{He}{2}\,$\lambda$4686 was present (a sufficient, though not necessary, criterion), the object was point-like, and no continuum was observed.  Based upon these criteria, S33, pn-010, pn-012, S16, and pn-017 appear to be planetary nebulae, in agreement with previous studies \citep{dufourtalent1980, richermccall1995, leisyetal2005a}.  At the distance of NGC 6822 \citep[0.5\,Mpc;][]{mateo1998}, a planetary nebula with a typical radius of 0.1\,pc should be unresolved.  KD29, Ho06, Ho10, and S10 are not point sources and therefore \ion{H}{2} regions.  

\begin{figure*}
\begin{center}
\includegraphics[width=0.4\linewidth]{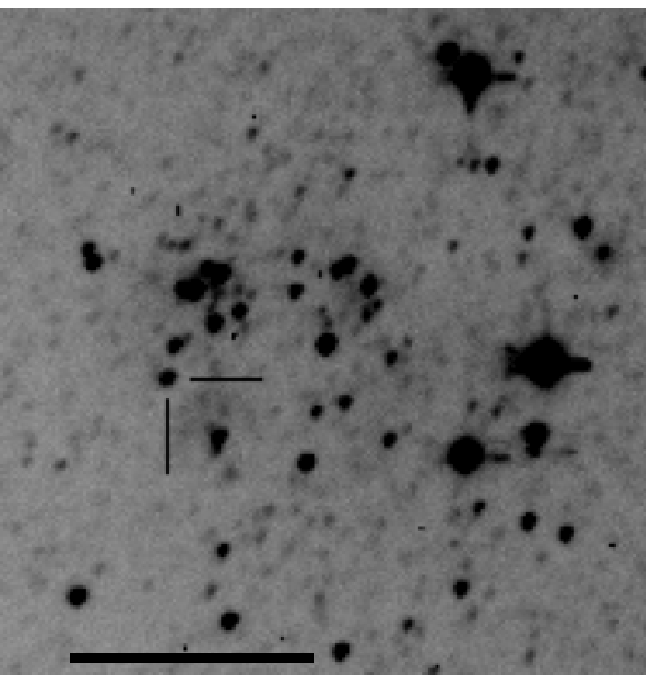}
\includegraphics[width=0.4\linewidth]{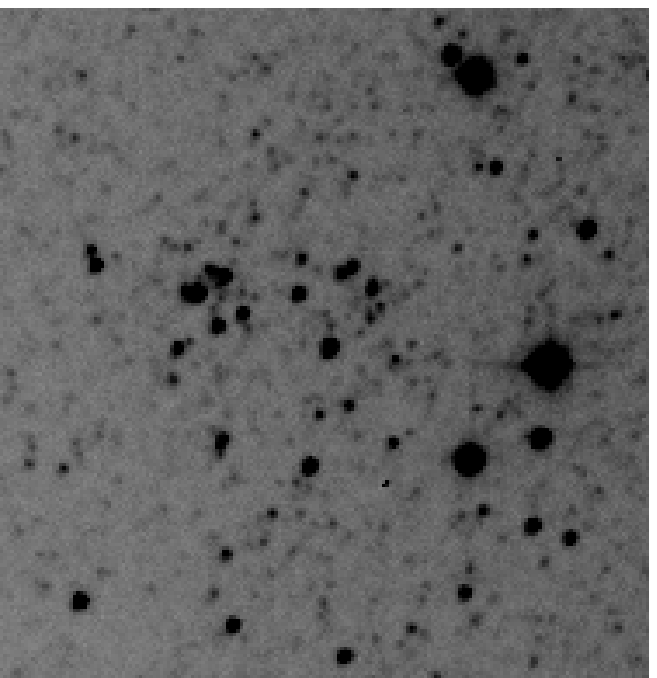}
\includegraphics[width=0.137\linewidth]{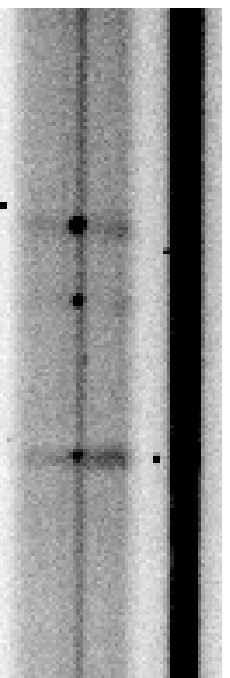}
\end{center}
\caption{From left to right, we present the on-band, off-band, and the two-dimensional spectrum of pn-020 (the wavelength interval including H$\beta$ at bottom and [\ion{O}{3}]$\lambda\lambda$4959,5007 at top; from UT 19 August 1998).  In the images, north is up and east is to the left.  In the spectrum, the spatial axis has east to the left.  The vertical and horizontal lines in the on-band image locate pn-020.  Its coordinates are $\alpha =$\,19:45:11.5 $\delta =$\,-14:48:54 (J2000, uncertainty $\pm 1^{\prime\prime}$).  The black bar at the bottom of the on-band image is 30$\arcsec$ long.  The spectrum clearly demonstrates that the planetary nebula is offset to the east from the faint star visible in the off-band image.  Likewise, the spectrum emphasizes the dramatically different line intensity ratios compared to the background \ion{H}{2} region Ho15 \label{fig_pn020}}
\end{figure*}

The foregoing leaves only pn-019 and pn-020 unclassified.  Both of these objects are point sources in our images, but neither present \ion{He}{2}\,$\lambda$4686 emission and both also present a continuum.  It is tempting to automatically disqualify any object with an observable continuum as a planetary nebula, but an obvious problem with this approach is that there are an abundance of bright stars in star-forming galaxies that chance may throw along the line of sight.  The case of pn-019 is the simpler of the two.  It is evident from the on- and off-band images that there is a star about 0\farcs 9 to the south that is also visible on the finding chart presented by \citet{leisyetal2005a}.  In all likelihood, it is this object that contributes the continuum emission.  In the case of pn-020, on- and off-band images as well as a section of the two-dimensional spectrum are presented in Fig. \ref{fig_pn020}.  In the two-dimensional spectrum, the line emission is displaced from the continuum by about 0\farcs 5 to the east.  A second, fainter continuum source appears about 1\farcs 5 farther to the east of the pn-020.  Careful inspection of the on- and off-band images reveals these two objects to be stars.  Consequently, we conclude that pn-020 is not a source of continuum emission.  The emission line spectrum further argues that this object is not an \ion{H}{2} region or an unresolved nova or supernova remnant:  the observed [\ion{O}{3}]$\lambda5007/\mathrm H\beta$ intensity ratio (Table \ref{table_line_int}) is nearly double the highest value observed in any \ion{H}{2} region in NGC 6822 \citep[][this study]{pageletal1980, peimbertetal2005, leeetal2006} and the [\ion{S}{2}]$\lambda\lambda$6716,6731 lines in this object are weaker than those in the four \ion{H}{2} regions observed here.  

To summarize, we find that pn-019 and pn-020 both appear to be planetary nebulae.  Fig. \ref{fig_pn020} serves to locate pn-020.  \citet{leisyetal2005a} present finding charts for the other planetary nebulae.  We conclude that our sample includes seven planetary nebulae and four \ion{H}{2} regions.  

\subsection{Physical Conditions and Chemical Abundances}

Table \ref{table_chem_abun} presents the electron temperatures and densities as well as the ionic and elemental abundances derived for each object.  The atomic data employed for ions of N, O, and Ne are listed in Table \ref{table_atomic_data}.  For H$^{\circ}$ and He$^+$, the emissivities of \citet{storeyhummer1995} were used.  For He$^{\circ}$, we used an extended list of the emissivities from \citet{porteretal2005} that was kindly provided by R. Porter.  

The ionic abundances were derived using the SNAP software package \citep{krawchuketal1997}.  The elemental abundances were calculated from the ionic abundances using the ionization correction factors (ICFs) proposed by \citet{kingsburghbarlow1994}.  The uncertainties quoted for all quantities account for the uncertainties in the line intensities involved in their derivation, including the uncertainties in the reddening.  However, the quoted uncertainties in the elemental abundances do not include the uncertainties in the ICFs.  Note that no ICFs are involved in the $\mathrm N/\mathrm O$ and $\mathrm{Ne}/\mathrm O$ abundance ratios that are tabulated in Table \ref{table_chem_abun}.

In the calculation of the elemental abundances, we made an effort to adopt a scheme that treated all objects as homogeneously as possible.  Occasionally, this will result in less than optimal elemental abundances for some objects, but has the advantage of uniformity, i.e., the drawbacks are common to all objects.  For example, it is feasible to use singlet lines of \ion{He}{1} to compute $\mathrm{He}^+/\mathrm H$ in some objects, but not in all.  To avoid the risk of introducing spurious differences, we chose to use common lines for all objects.  The one exception to this rule is the $\mathrm O^+/\mathrm H$ ionic abundance.  When available, the ionic abundance based upon [\ion{O}{2}]$\lambda$3727 was used in favor of that based upon [\ion{O}{2}]$\lambda\lambda$7319,7331.  The electron temperature adopted is always based upon the [\ion{O}{3}]$\lambda\lambda$4363/5007 ratio, the upper limit being used when [\ion{O}{3}]$\lambda$4363 was not detected.  Likewise, the only electron densities adopted were based upon the [\ion{S}{2}]$\lambda\lambda$6716,6731 lines, substituting values of 100\,cm$^{-3}$ and 2000\,cm$^{-3}$ for \ion{H}{2} regions and planetary nebulae, respectively, when the density could not be derived.  The density adopted for planetary nebulae was chosen based upon the results of \citet{riesgolopez2006}.

Oxygen abundances for the planetary nebulae S16 and S33 were previously presented by \citet{dufourtalent1980} and \citet{richermccall1995}.  Within uncertainties, the oxygen abundances found here agree with their previous values.  We note that the electron temperatures we find for the planetary nebulae are generally higher than those found for the \ion{H}{2} regions, as expected.

Our oxygen abundances for the \ion{H}{2} regions KD29, Ho10, and Ho06 are similar to those derived previously for other \ion{H}{2} regions in NGC 6822 by \citet{pageletal1980}, \citet{peimbertetal2005}, and \citet{leeetal2006} for the objects in which they detect [\ion{O}{3}]$\lambda$4363.  For KD29, we only have an upper limit to [\ion{O}{3}]$\lambda$4363 (Table \ref{table_line_int}), but it is quite low, so there is reason to believe that the oxygen abundance we derive is not unreasonably low.  For S10, we have a reasonable detection of [\ion{O}{3}]$\lambda$4363, so its low oxygen abundance is somewhat surprising.  

Had we adopted lower temperatures for the O$^+$ zone, as done in \citet{leeetal2006}, we would find slightly higher abundances for all objects except Ho06 (its electron temperature would hardly change in the \citet{leeetal2006} scheme).  The reverse would be true had we adopted higher temperatures in the O$^+$ zone as \citet{peimbertetal2005} did.  \citet{leeetal2006} find a scatter in their measured oxygen abundances of about $\pm 0.1$\,dex.  We find about double this scatter, driven by the low abundance for S10.  
%On the other hand, the scatter we find is less than that found in the Milky Way disk \citep{boesgaard1989, rana1991}.  
The helium abundances we derive for the \ion{H}{2} regions in our sample are similar to those obtained by \citet{pageletal1980} and \citet{peimbertetal2005}.

\section{Chemical Similarity between Planetary Nebulae and \ion{H}{2} Regions}

We now consider the data available in the literature for bright planetary nebulae in dwarf irregular galaxies.  By bright planetary nebulae, we adopt all planetary nebulae within 2\,mag of the brightest in each galaxy.  Selecting planetary nebulae based upon high [\ion{O}{3}]$\lambda$5007 luminosity  affects the resulting sample in two known ways (and perhaps others so far unidentified).  First, high [\ion{O}{3}]$\lambda$5007 luminosity will favor the most oxygen-rich planetary nebulae in each galaxy, at least for oxygen abundances below that of the ISM in the SMC \citep{dopitaetal1992, richermccall1995}.  As we argue below, this should favor objects resulting from recent star formation.  Second, high [\ion{O}{3}]$\lambda$5007 luminosity will also favor planetary nebulae early in their evolution \citep[e.g., ][]{jacoby1989, stasinskaetal1998}.  

In Table \ref{table_di_prop}, we compare the average oxygen abundances in the bright planetary nebulae in dwarf irregulars with the corresponding average oxygen abundances in their \ion{H}{2} regions.  For all of the planetary nebulae in all of the dwarf irregular galaxies, we have re-computed their chemical abundances from the line intensities given in the data sources, following the procedure outlined above for the planetary nebulae in NGC 6822.  For the \ion{H}{2} regions in the Magellanic Clouds, Sextans A and B, and Leo A, we have adopted the abundances given in the original studies.  

Figure \ref{fig_oh_gap} presents the information in Table \ref{table_di_prop} graphically.  In all of the galaxies considered, the mean oxygen abundance in planetary nebulae and \ion{H}{2} regions is the same, within uncertainties, except in Sextans A.  If Leo A is excluded, the data suggest a trend.  However, the significance of the slope of a line fit to the data (excluding Leo A) is then critically dependent upon the inclusion of Sextans A, whose \lq\lq population" of bright planetary nebulae contains a single member.  If Sextans A is excluded, an F-test indicates that the slope of the resulting relation is not statistically significant.  If Sextans A is included, the slope of the line fit to the data is similar, but its statistical significance improves notably, reducing the probability of obtaining the relation by chance to 6\%.  Independent of the existence of any trend, Fig. \ref{fig_oh_gap} indicates that the dredge-up of oxygen is uncommon among the progenitors of the bright planetary nebulae, since the mean abundance found for bright planetary nebulae agrees with that in the interstellar medium, at least for dwarf irregular galaxies with metallicities similar to or greater than that of the SMC.
%Unfortunately, when only one object is available, it is difficult to judge whether it is representative.  It is not clear from this plot how deviant Sextans A is, since its population of planetary nebulae comprises one object.  The solid line in Fig. \ref{fig_oh_gap} is a fit to the points for the Magellanic Clouds, Sextans B, and NGC 6822.  Unfortunately, given the example of the planetary nebula in Leo A, it is difficult to judge how representative the planetary nebulae in Sextans A.  

\begin{figure*}
\includegraphics[angle=90,scale=0.56]{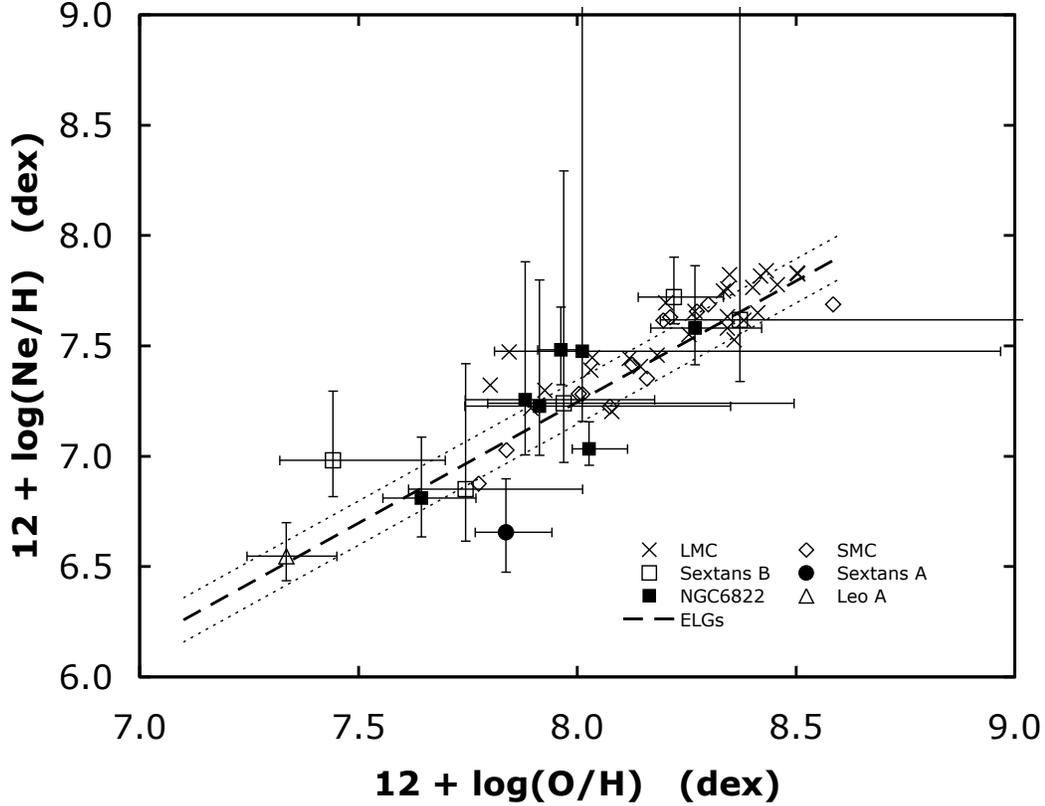}
\caption{We present the correlation of neon and oxygen abundances in bright planetary nebulae in the sample of dwarf irregular galaxies for which these data exist.  In this and subsequent plots, no uncertainties are shown for the planetary nebulae in the LMC and SMC since this information is not available from the original sources, though it is likely that their uncertainties are similar to those for the objects in other galaxies.  The heavy dashed line is the relation between neon and oxygen abundances in \ion{H}{2} regions in emission line galaxies (ELGs) from \citet{izotovetal2006}, while the thin dashed lines indicate the scatter about this relation.  The generally excellent agreement between the planetary nebulae and the ELGs indicates that the stellar progenitors of most bright planetary nebulae in dwarf irregulars do not significantly modify either of these abundances.\label{fig_ne_o_pn}}
\end{figure*}

Figure \ref{fig_ne_o_pn} illustrates the excellent correlation between neon and oxygen abundances in bright planetary nebulae in dwarf irregular galaxies.  No error bars are drawn for the planetary nebulae in the Magellanic Clouds since this information is not generally available from the original sources.  However, it is likely that the uncertainties are similar to those for the other planetary nebulae.  In this figure, we also plot the relationship between these abundances found by \citet{izotovetal2006} in emission line galaxies, where the relation is set by the nucleosynthetic yields of type II supernovae.  The agreement between the two relations is excellent.  A linear least squares fit to the planetary nebula data yields
\begin{equation}
12+\log(\mathrm{Ne}/\mathrm H) = (1.043\pm 0.075)X - 1.04\pm 0.61
\end{equation}

\noindent where $X = 12+\log(\mathrm O/\mathrm H)$.  Within uncertainties, the slope and intercept coincide with the relation found by \citet{izotovetal2006}.  Since the stellar progenitors of planetary nebulae are not expected to modify their neon abundance during their evolution \citep{marigoetal2003}, the good correlation found between the oxygen and neon abundances indicates that the progenitors of bright planetary nebulae in dwarf irregulars generally do not modify either abundance.  This correlation is the basis for supposing that the most oxygen-rich planetary nebulae in dwarf irregulars are the result of recent star formation rather than dredge-up.

\begin{figure*}
\includegraphics[angle=90,scale=0.56]{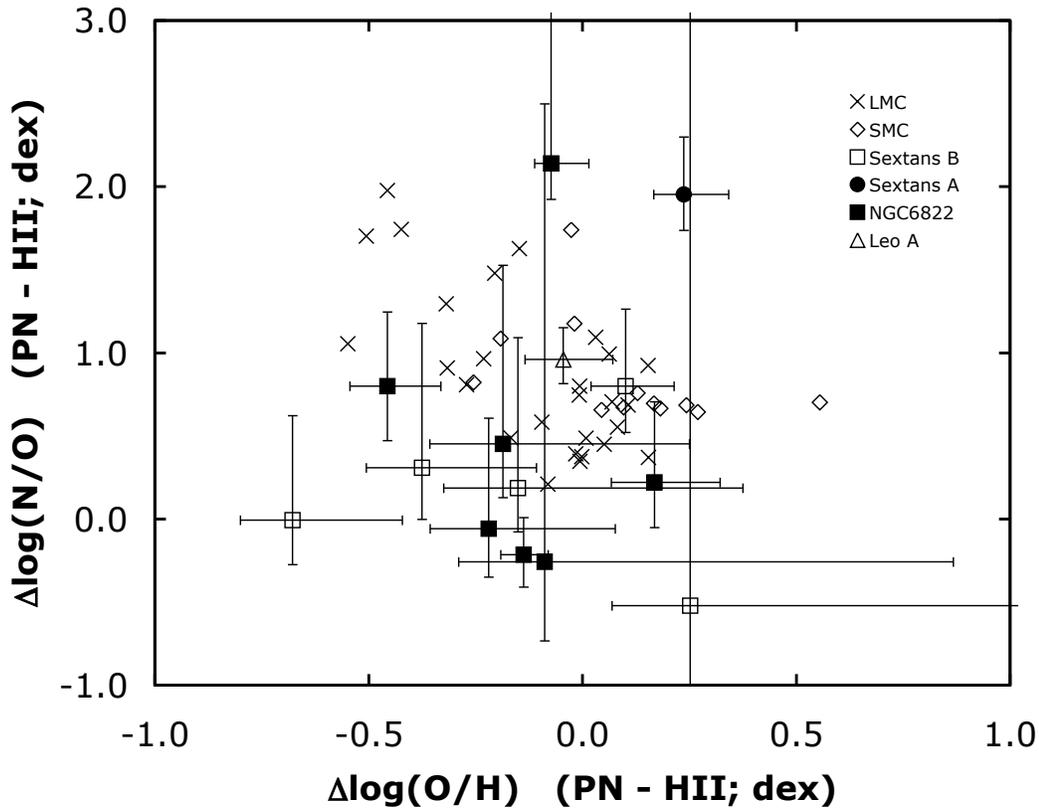}
\caption{We plot the $\mathrm N/\mathrm O$ abundance ratio as a function of the oxygen abundance for bright planetary nebulae in dwarf irregular galaxies.  Both $\log(\mathrm N/\mathrm O)$ and $\log(\mathrm O/\mathrm H)$ are computed with respect to the values found in the ISM for the host galaxies in order to emphasize that most bright planetary nebulae have oxygen abundances similar to those in the ISM and that sometimes even nitrogen is enriched very little.  \label{fig_diff_n_o}}
\end{figure*}

Figure \ref{fig_diff_n_o} plots the $\mathrm N/\mathrm O$ abundance ratio as a function of the oxygen abundance.  In this figure, both the $\mathrm N/\mathrm O$ ratio and the oxygen abundance in the planetary nebulae are calculated relative to the values observed in the ISM (\ion{H}{2} regions).  This figure emphasizes that most of the bright planetary nebulae in these galaxies have oxygen abundances close to the ISM values.  It also emphasizes how little $\mathrm N/\mathrm O$ varies from its ISM values in some of these galaxies, particularly NGC 6822 and Sextans B.  

\section{Nucleosynthesis in the Progenitors of Planetary Nebulae}

\begin{figure*}
\includegraphics[angle=-90,scale=0.56]{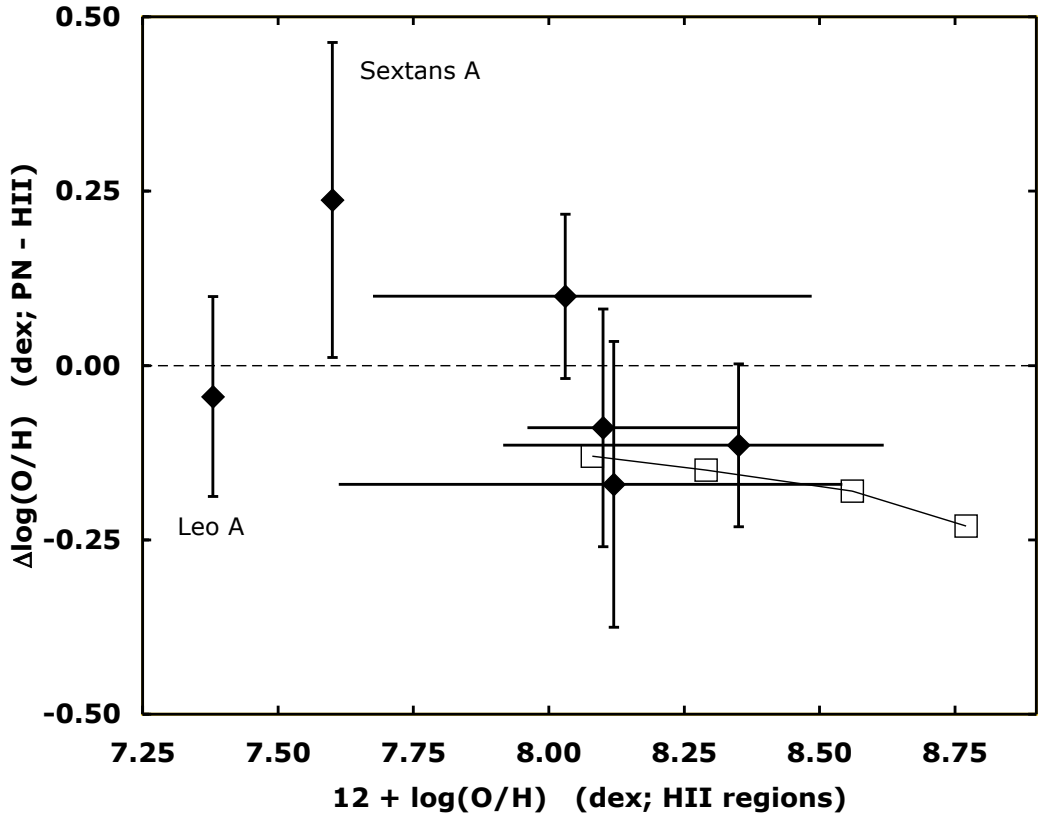}
\caption{We compare the mean oxygen abundance in bright planetary nebulae and \ion{H}{2} regions in dwarf irregular galaxies.  The vertical error bars indicate the uncertainty in the ratio of abundances.  The horizontal bar on each point indicates the abundance range spanned by the planetary nebulae in each galaxy.  No range is shown for Leo A or Sextans A since only one planetary nebula is known in each of these galaxies.  The horizontal dashed line indicates agreement between oxygen abundances in planetary nebulae and \ion{H}{2} regions.  Generally, the oxygen abundances in bright planetary nebulae are very similar to those in \ion{H}{2} regions.  The open squares connected by a line are the predictions for star-forming galaxies from \citet{richeretal1997}.  \label{fig_oh_gap}}
\end{figure*}

Returning to Fig. \ref{fig_oh_gap}, population synthesis models indicate that the oxygen abundance in the planetary nebula population is always expected to be below that observed in the interstellar medium \citep{richeretal1997}, though those models assume that the progenitors of planetary nebulae do not dredge up oxygen.  As the oxygen abundance decreases, these models indicate that the difference in abundance between the planetary nebulae and the interstellar medium should decrease.  The sense of any possible trend in Fig. \ref{fig_oh_gap} is therefore in agreement with these expectations.  The novelty is that Fig. \ref{fig_oh_gap} indicates that the bright planetary nebulae may in fact have oxygen abundances exceeding those in the interstellar medium at low oxygen abundances.  In the context of population synthesis models, an oxygen abundance in planetary nebulae exceeding that found in the interstellar medium requires that the progenitors of the planetary nebulae commonly dredge up oxygen. 

%The above conclusion actually underestimates the situation somewhat.  Although the planetary nebulae in dwarf irregulars tend to have oxygen abundances very similar to those currently existing in the interstellar medium, as Fig. \ref{fig_diff_n_o} demonstrates, some of them will be products of star formation some time in the past.  

\begin{figure*}
\includegraphics[angle=-90,scale=0.56]{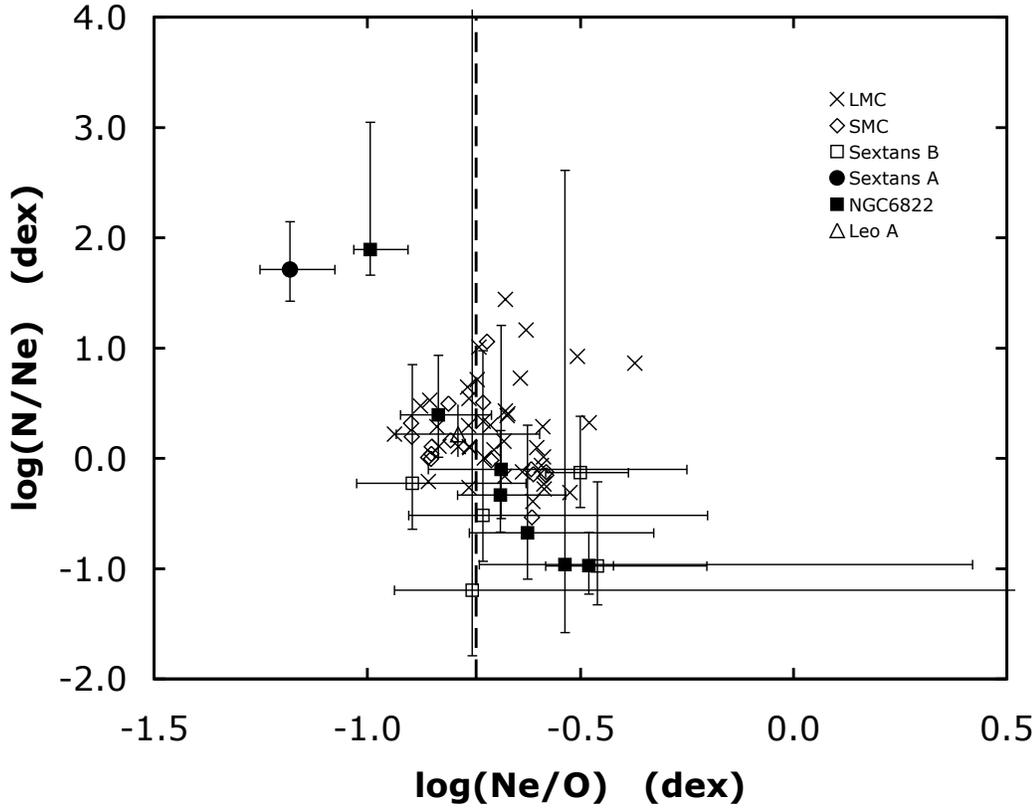}
\caption{Here, we plot the $\mathrm N/\mathrm{Ne}$ abundance ratio as a function of the $\mathrm{Ne}/\mathrm O$ abundance ratio for bright planetary nebulae in dwarf irregular galaxies.  Objects that have dredged up oxygen should have low values of $\mathrm{Ne}/\mathrm O$, such as occurs for the planetary nebula in Sextans A and S33 in NGC 6822, while $\mathrm N/\mathrm{Ne}$ may vary according to which dredge-up episodes have occurred.  The vertical dashed line is the value of the $\mathrm{Ne}/\mathrm O$ ratio in star-forming galaxies for $12+\log(\mathrm O/\mathrm H) = 8.0$\,dex \citep{izotovetal2006}. \label{fig_n_o_ne}}
\end{figure*}

The planetary nebula in Sextans A has a strong bearing on whether oxygen is judged to be dredged up at low metallicity.  In Fig. \ref{fig_n_o_ne}, we present the $\mathrm N/\mathrm{Ne}$ abundance ratio as a function of the $\mathrm{Ne}/\mathrm O$ abundance ratio for the planetary nebulae in our sample.  We adopt these abundance ratios since neon is neither expected nor observed to change significantly as a result of the prior evolution of the progenitors of planetary nebulae \citep{marigoetal2003}.  We also plot the $\mathrm{Ne}/\mathrm O$ ratio found at an oxygen abundance of $12+\log(\mathrm O/\mathrm H)=8.0$\,dex \citep{izotovetal2006}.  If oxygen is dredged up, one expects that $\mathrm{Ne}/\mathrm O$ should be low.  $\mathrm N/\mathrm{Ne}$ will depend upon which processes have dredged up nitrogen \citep{marigo2001}.  Low $\mathrm{Ne}/\mathrm O$ is found for the planetary nebula in Sextans A and for S33 in NGC 6822.  Thus, oxygen is dredged up on occasion.  

While the above examples support the hypothesis that oxygen can occasionally be dredged up, they do not address the principal concern, which is whether oxygen is routinely dredged up at low metallicity.  The planetary nebula in Leo A provides useful guidance.  As \citet{vanzeeetal2006} argue, this object may not be the result of recent star formation in Leo A, but may have been formed some time in the past.  
%Given the very low oxygen abundance in Leo A, that the oxygen abundance found in this planetary nebula agrees with that found in the interstellar medium might simply reflect the primitive chemical state of the galaxy.  
The oxygen abundance in this object is the same as that in the interstellar medium.  Furthermore, the nebula falls on the neon-oxygen relation for star-forming galaxies.  It follows that the progenitor of this planetary nebula did not substantially modify its initial store of oxygen.  If one accepts that the planetary nebula in Leo A is not derived from a recently formed, and therefore relatively massive, stellar progenitor, the comparison of this object with the planetary nebula in Sextans A allows the possibility that only massive progenitors of planetary nebulae dredge up oxygen, provided, of course, that the planetary nebula in Sextans A descends from a massive progenitor.  If the planetary nebula in Leo A is the result of recent star formation, one must conclude that oxygen is only dredged up under some fraction of circumstances, even at very low metallicity.

The prevalence of oxygen self-enrichment should become clearer in the near future.  Three groups have undertaken spectroscopy of planetary nebulae in NGC 3109 \citep{leisyetal2005b, kniazevetal2006, penaetal2006}.  Given NGC 3109's oxygen abundance of $12+\log(\mathrm O/\mathrm H)\sim 7.8$\,dex \citep{leeetal2003}, its population of planetary nebulae should help clarify whether the trend in Fig. \ref{fig_oh_gap} is significant.  Likewise, spectroscopy of the planetary nebulae recently reported in IC 1613 and WLM would be helpful \citep[both galaxies have $12+\log(\mathrm O/\mathrm H)\sim 7.7$\,dex; ][]{leeetal2003}.  It would also be useful to confirm that planetary nebulae have systematically lower abundances than the interstellar medium at oxygen abundances exceeding $12+\log(\mathrm O/\mathrm H)=8$\,dex.  Spectroscopy of the planetary nebulae in IC 10 \citep[$12+\log(\mathrm O/\mathrm H)\sim 8.2$\,dex; ][]{leeetal2003} or in the disks of M31, M33, or other nearby spirals would be useful in this regard.  (Planetary nebulae in the disk of the Milky Way are less desirable since their absolute luminosities are much more difficult to ascertain.)  

\begin{figure*}
\includegraphics[angle=90,scale=0.56]{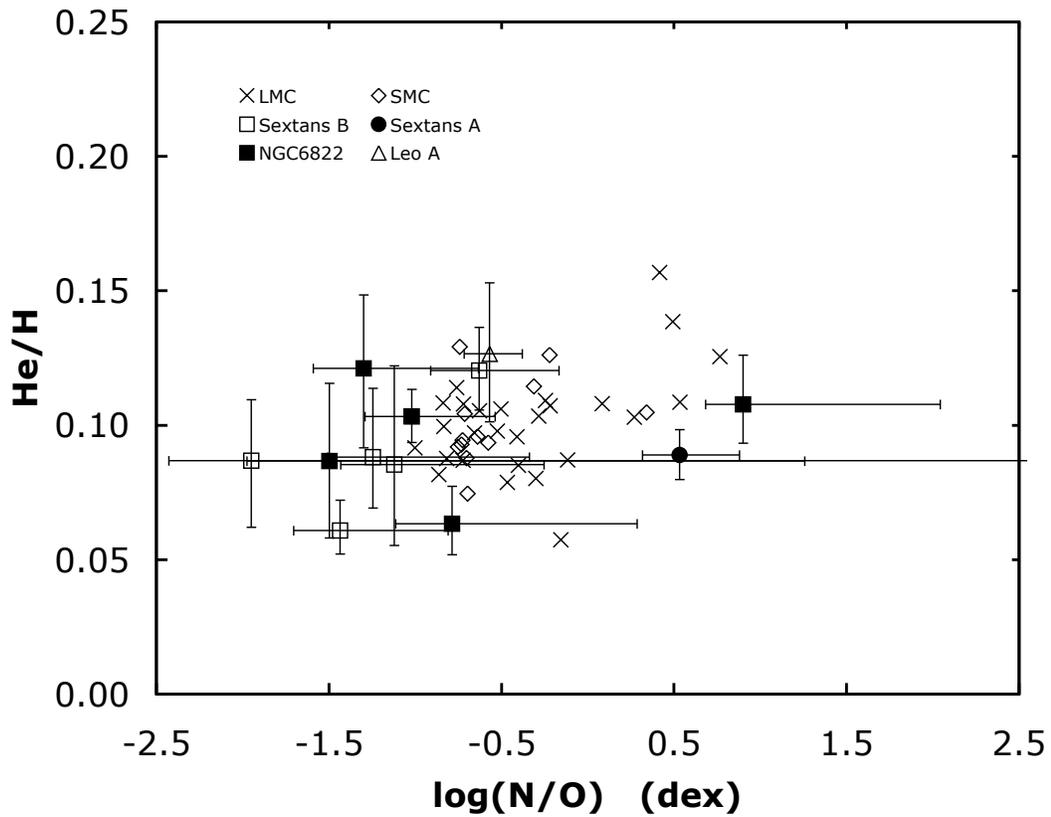}
\caption{Here, we plot the helium abundance as a function of the $\mathrm{N}/\mathrm O$ abundance ratio for bright planetary nebulae in dwarf irregular galaxies.  The relation is surprisingly flat.  It is also perhaps noteworthy that a similar limiting nitrogen enrichment, $\log(\mathrm N/\mathrm O)\sim 0.5$\,dex, is achieved in most galaxies, independently of the helium abundance. \label{fig_he_no}}
\end{figure*}

Figure \ref{fig_diff_n_o} emphasizes the difficulty of interpreting nitrogen abundances.  It is apparent that $\mathrm N/\mathrm O$ does not vary in a uniform way as a function of oxygen abundance.  In the LMC, the planetary nebulae with the highest oxygen abundances have the lowest $\mathrm N/\mathrm O$ ratios.  As the oxygen abundance decreases, the $\mathrm N/\mathrm O$ ratio increases, indicating that the progenitors of these objects enriched themselves more efficiently in nitrogen.  In contrast, the planetary nebulae in the SMC are all approximately uniformly enriched in N compared to the ISM by $\Delta\log(\mathrm{N}/\mathrm O)\sim 0.7$\,dex.  For the other galaxies, no clear trends are evident, though it is noteworthy that the bright planetary in most of them achieve approximately the same limiting nitrogen enrichment relative to oxygen, $\log(\mathrm N/\mathrm O)\sim 0.5$\,dex (cf. Fig. \ref{fig_he_no}).  Perhaps, the simplest conclusion is that the progenitors of all planetary nebulae enrich themselves in nitrogen to some arbitrary extent, depending upon parameters other than initial mass, as appears to occur in the progenitors of planetary nebulae in systems without star formation \citep{richer2006}.

An important question is whether nitrogen enrichment occurs as a result of the consumption of carbon or of oxygen also.  (The conversion of carbon to nitrogen should occur in any case.)  That the neon and oxygen abundances for the planetary nebulae in dwarf irregulars generally follow the trend found in ELGs argues that the nitrogen enrichment comes at the expense of carbon.  (The same conclusion is drawn from plotting the $\mathrm{Ne}/\mathrm O$ abundance ratio as a function of oxygen abundance.)  

The $\mathrm{N}/\mathrm O$ ratio may reflect the chemical evolution of the host galaxies as well as the nucleosynthesis within the progenitors of these planetary nebulae.  Many of the planetary nebulae in Sextans B and NGC 6822 are examples of the interplay of these influences since they generally have both oxygen abundances and $\mathrm{N}/\mathrm O$ ratios that agree with the value found in the interstellar medium.  Either their progenitors did not dredge up nitrogen, which is surprising if they were relatively massive, or these galaxies have evolved chemically very little in the recent past, allowing the progenitors of their bright planetary nebulae to be less massive.  The latter option, however, is somewhat at odds with the expectation that bright planetary nebulae should arise from progenitors of order 2\,$M_{\sun}$ \citep[e.g., ][]{marigoetal2004}.  The lack of nitrogen production is surprising, especially if the stellar progenitors are massive, since both the first and second dredge-up as well as hot bottom burning should all result in nitrogen production \citep[e.g., ][]{forestinicharbonnel1997}.  At any rate, if nitrogen, which should be more easily produced, is not always dredged up in these progenitors of planetary nebulae, oxygen should be dredged up even less frequently.  

Finally, in Fig. \ref{fig_he_no}, we plot the helium abundances for the planetary nebulae in our sample as a function of $\log(\mathrm N/\mathrm O)$.  There is little correlation between helium abundance and $\log(\mathrm N/\mathrm O)$.  Comparing the helium abundances in planetary nebulae and \ion{H}{2} regions in NGC 6822 (Table \ref{table_chem_abun}), the helium enrichment in planetary nebulae spans the range from zero to a doubling of the original helium content.  

\section{Conclusions}

We have obtained spectroscopy of a sample of seven planetary nebulae and four \ion{H}{2} regions in the Local Group dwarf irregular galaxy NGC 6822.  From these data, we calculate their chemical composition and find that the planetary nebulae in NGC 6822 have oxygen and neon abundances very similar to those in the interstellar medium, from which we infer that these planetary nebulae arise preferentially, though not exclusively, as a result of recent star formation in NGC 6822.  
%Considering similar data available in the literature for planetary nebulae in other dwarf irregular galaxies, we arrive at a similar conclusion concerning the origin of their bright planetary nebulae.  

The difference between the mean oxygen abundance in bright planetary nebulae and the mean for HII regions is consistent with zero regardless of the level of enrichment, except perhaps in Sextans A, but the result for this galaxy is based upon a single planetary nebula.  Very generally, we find that bright planetary nebulae in dwarf irregulars have oxygen and neon abundances similar to those found in the interstellar medium in star-forming galaxies.  Since the latter abundance ratio is set by the nucleosynthetic yields of type II supernovae, and since the progenitors of planetary nebulae are not expected to modify their neon abundances, it is unlikely that the progenitors of the vast majority of the bright planetary nebulae in our sample modified their initial oxygen or neon abundances.  

Two planetary nebulae in our combined sample, that in Sextans A and S33 in NGC 6822, do present abundance ratios indicative of the dredge-up of oxygen.  While it remains to be shown that oxygen is routinely dredged up, there is a slight  suggestion of a trend wherein this might happen at low metallicity under some circumstances.  On the other hand, many of the progenitors of the bright planetary nebulae in Sextans B and NGC 6822 did not dredge up nitrogen, implying that the dredge-up of nitrogen might be a less common process than has been thought.  

\acknowledgments

We thank the time allocation committee of the Canada-France-Hawaii Telescope for granting us the opportunity to observe.  We thank R. Porter for providing us with a more extensive list of He$^{\circ}$ emissivities.  MGR acknowledges financial support from CONACyT through grant 43121 and from UNAM-DGAPA via grants IN112103, IN108406-2, and IN108506-2.  MLM thanks the Natural Sciences and Engineering Research Council of Canada for its continuing support.

\facility{Canada-France-Hawaii Telescope (Multi-Object Spectrograph)}

\clearpage

\begin{deluxetable}{llccccccc}
\tabletypesize{\scriptsize}
%\rotate
\tablecaption{Observed Line Intensities$^{\mathrm a}$ for PNe$^{\mathrm b}$ and \ion{H}{2} regions$^{\mathrm b}$ in NGC 6822\label{table_line_int}}
\tablewidth{0pt}
%\tablehead{
%\colhead{wavelength} & \colhead{ion} & \colhead{pn-020} & \colhead{S33} & \colhead{pn-010} &
%\colhead{pn-012} & \colhead{S16} & \colhead{pn-017} &
%}
\startdata
\hline\hline\\[1pt]
wavelength & ion & $A_1(\lambda)^{\mathrm c}$ & pn-020  & S33        & pn-010           & pn-012           & S16               & pn-017 \\[3pt]
\hline\\[1pt]
3727       & [O II]   & 4.52 & $ <51           $ & $  28.4\pm 7.2  $ & $318\pm 51     $ & $ 92\pm 42     $ & $  22           $ & $              $ \\
3868.76    & [Ne III] & 4.39 & $  41\pm 11     $ & $  31.7\pm 2.4  $ & $ 52\pm 14     $ & $ 50\pm 18     $ & $  46.5\pm 6.5  $ & $<47           $ \\
3889.049   & H I      & 4.37 & $               $ & $   9.8\pm 2.0  $ & $              $ & $              $ & $               $ & $              $ \\
3967.47    & [Ne III] & 4.30 & $               $ & $  10.6\pm 2.2  $ & $              $ & $              $ & $   8.5\pm 5.8  $ & $ 49\pm 15     $ \\
3970.072   & H I      & 4.29 & $               $ & $   8.9\pm 2.2  $ & $              $ & $              $ & $   9.3\pm 5.9  $ & $              $ \\
4101.765   & H I      & 4.18 & $               $ & $  21.3\pm 1.8  $ & $              $ & $              $ & $  15.0\pm 3.2  $ & $<34           $ \\
4340.495   & H I      & 3.97 & $ <11           $ & $  43.6\pm 1.4  $ & $ 34.2\pm 7.8  $ & $ 49.2\pm 8.7  $ & $  34.9\pm 3.6  $ & $ 37.5\pm 9.2  $ \\
4363.21    & [O III]  & 3.95 & $ <11           $ & $  29.1\pm 1.2  $ & $ 20.2\pm 6.3  $ & $ 10.9\pm 6.8  $ & $  14.0\pm 2.9  $ & $ 23.4\pm 8.3  $ \\
4387.929   & He I     & 3.93 & $               $ & $   2.2\pm 0.9  $ & $              $ & $              $ & $               $ & $              $ \\
4471.477   & He I     & 3.86 & $               $ & $   3.43\pm 0.97$ & $              $ & $              $ & $               $ & $              $ \\
4685.75    & He II    & 3.66 & $  <8           $ & $  56.7\pm 1.3  $ & $ 35.4\pm 6.5  $ & $<13           $ & $  20.9\pm 2.2  $ & $ 41.4\pm 9.1  $ \\
4711.34    & [Ar IV]  & 3.63 & $               $ & $   0.4\pm 1.1  $ & $              $ & $              $ & $               $ & $              $ \\
4713.375   & He I     & 3.63 & $               $ & $   1.2\pm 1.2  $ & $              $ & $              $ & $               $ & $              $ \\
4740.2     & [Ar IV]  & 3.62 & $               $ & $   3.76\pm 0.81$ & $              $ & $              $ & $               $ & $              $ \\
4861.332   & H I      & 3.49 & $ 100.0\pm 4.9  $ & $ 100.0\pm 1.0  $ & $100.0\pm 4.7  $ & $100.0\pm 6.2  $ & $ 100.0\pm 2.1  $ & $100.0\pm 7.6  $ \\
4958.92    & [O III]  & 3.39 & $ 323\pm 19     $ & $ 380.5\pm 7.2  $ & $265\pm 15     $ & $257\pm 18     $ & $ 424\pm 14     $ & $239\pm 22     $ \\
5006.85    & [O III]  & 3.34 & $1009\pm 51     $ & $1178.0\pm 14   $ & $787\pm 38     $ & $789\pm 50     $ & $1310\pm 30     $ & $747\pm 59     $ \\
5200       & [N I]    & 3.19 & $               $ & $   9.5\pm 1.1  $ & $              $ & $              $ & $               $ & $              $ \\
5412       & He II    & 3.02 & $               $ & $   6.33\pm 0.82$ & $              $ & $              $ & $               $ & $              $ \\
5754.57    & [N II]   & 2.75 & $               $ & $  30.3\pm 1.4  $ & $              $ & $              $ & $               $ & $              $ \\
5875.666   & He I     & 2.65 & $  33.6\pm 5.3  $ & $  13.9\pm 1.2  $ & $ 16.0\pm 5.7  $ & $ 17.6\pm 6.7  $ & $  24.8\pm 2.6  $ & $ <6           $ \\
6300.32    & [O I]    & 2.40 & $               $ & $  40.1\pm 1.5  $ & $              $ & $              $ & $               $ & $              $ \\
6312.06    & [S III]  & 2.39 & $               $ & $   1.5\pm 1.0  $ & $              $ & $              $ & $               $ & $              $ \\
6363.81    & [O I]    & 2.36 & $               $ & $  15.3\pm 1.3  $ & $              $ & $              $ & $               $ & $              $ \\
6548.06    & [N II]   & 2.26 & $               $ & $ 243.3\pm 5.3  $ & $ 12.0\pm 6.0  $ & $ 22.5\pm 8.3  $ & $  20.5\pm 9.1  $ & $ 31.3\pm 9.7  $ \\
6562.817   & H I      & 2.26 & $1097\pm 55     $ & $ 477.4\pm 7.0  $ & $455\pm 23     $ & $545\pm 36     $ & $ 703\pm 19     $ & $420\pm 34     $ \\
6583.39    & [N II]   & 2.25 & $ <12           $ & $ 720.0\pm 9.2  $ & $ 34.4\pm 7.1  $ & $ 12.5\pm 7.8  $ & $  36.2\pm 9.4  $ & $ 60\pm 11     $ \\
6678.149   & He I     & 2.20 & $  12.5\pm 3.5  $ & $   3.66\pm 0.97$ & $  6.4\pm 5.3  $ & $ <9           $ & $  14.0\pm 3.6  $ & $              $ \\
6716.42    & [S II]   & 2.17 & $   4.4\pm 3.2  $ & $   4.39\pm 0.99$ & $ 77.0\pm 9.0  $ & $ <9           $ & $  13.4\pm 3.5  $ & $ 40.2\pm 9.5  $ \\
6730.78    & [S II]   & 2.16 & $  20.2\pm 4.3  $ & $   5.8\pm 1.1  $ & $ 61.8\pm 8.4  $ & $ <9           $ & $   3.5\pm 2.6  $ & $ 19.8\pm 7.4  $ \\
7065.179   & He I     & 2.01 & $  46.3\pm 8.7  $ & $   8.6\pm 1.5  $ & $              $ & $              $ & $  26.9\pm 3.4  $ & $              $ \\
7135.8     & [Ar III] & 1.98 & $  57.3\pm 9.2  $ & $   7.9\pm 1.5  $ & $              $ & $              $ & $  17.9\pm 3.1  $ & $              $ \\
7319.92    & [O II]   & 1.90 & $  27.8\pm 8.0  $ & $               $ & $              $ & $<26           $ & $  13.7\pm 3.3  $ & $<25           $ \\
7330.19    & [O II]   & 1.90 & $   9.6\pm 6.1  $ & $               $ & $              $ & $<26           $ & $  12.8\pm 3.3  $ & $<25           $ \\
$E(B-V)_{\mathrm H\alpha}$ & (mag) && $ 1.21\pm 0.05$ & $ 0.49\pm 0.02$ & $ 0.45\pm 0.06$ & $ 0.59\pm 0.08$ & $ 0.82\pm 0.03$ & $  0.38\pm 0.08$ \\
$E(B-V)_{\mathrm H\gamma}$ & (mag) && $         $ & $   0.19\pm 0.07$ & $  0.74\pm 0.53$ & $ -0.09\pm 0.41$ & $ 0.68\pm 0.23$ & $  0.53\pm 0.57$ \\
\hline
\tablebreak
\hline\hline\\[1pt]
wavelength & ion      & $A_1(\lambda)^{\mathrm c}$ & pn-019            & S10               & KD29             & Ho10             & Ho06        \\[3pt]
\hline\\[1pt]
3727       & [O II]   & 4.52 & $  <6           $ & $               $ & $358\pm 15     $ & $              $ & $ 162\pm 12     $ \\
3868.76    & [Ne III] & 4.39 & $  19.7\pm 4.4  $ & $   8.1\pm 2.2  $ & $              $ & $              $ & $               $ \\
3889.049   & H I      & 4.37 & $  11.0\pm 3.6  $ & $   7.7\pm 1.9  $ & $              $ & $              $ & $               $ \\
3967.47    & [Ne III] & 4.30 & $   1.8\pm 2.3  $ & $               $ & $              $ & $              $ & $               $ \\
3970.072   & H I      & 4.29 & $   5.4\pm 2.5  $ & $   4.8\pm 1.6  $ & $              $ & $              $ & $               $ \\
4101.765   & H I      & 4.18 & $  17.1\pm 2.3  $ & $  12.4\pm 1.4  $ & $ 17.2\pm 3.5  $ & $              $ & $  19.9\pm 2.1  $ \\
4340.495   & H I      & 3.97 & $  41.2\pm 1.8  $ & $  34.0\pm 1.6  $ & $ 44.9\pm 3.4  $ & $ 31.9\pm 5.0  $ & $  41.3\pm 1.8  $ \\
4363.21    & [O III]  & 3.95 & $   8.5\pm 1.4  $ & $   3.9\pm 1.1  $ & $  1.4\pm 1.7  $ & $  5.0\pm 3.6  $ & $   2.9\pm 1.2  $ \\
4471.477   & He I     & 3.86 & $   4.5\pm 1.3  $ & $               $ & $  4.1\pm 2.3  $ & $              $ & $               $ \\
4685.75    & He II    & 3.66 & $  <2           $ & $               $ & $              $ & $              $ & $               $ \\
4861.332   & H I      & 3.49 & $ 100.0\pm 1.6  $ & $ 100.0\pm 2.0  $ & $100.0\pm 1.6  $ & $100.0\pm 3.2  $ & $ 100.00\pm 0.96$ \\
4958.92    & [O III]  & 3.39 & $ 159.5\pm 5.4  $ & $  94.6\pm 4.9  $ & $ 36.7\pm 1.9  $ & $146\pm 10     $ & $ 142.8\pm 2.3  $ \\
5006.85    & [O III]  & 3.34 & $ 487.6\pm 9.6  $ & $ 302.5\pm 8.1  $ & $109.6\pm 2.8  $ & $436\pm 18     $ & $ 442.9\pm 4.7  $ \\
5200       & [N I]    & 3.19 & $               $ & $   2.40\pm 0.68$ & $              $ & $              $ & $               $ \\
5875.666   & He I     & 2.65 & $  29.1\pm 1.9  $ & $  10.71\pm 0.91$ & $ 13.4\pm 1.7  $ & $ 12.4\pm 2.2  $ & $  12.7\pm 1.3  $ \\
6300.32    & [O I]    & 2.40 & $               $ & $  12.2\pm 1.2  $ & $              $ & $ 15.4\pm 2.3  $ & $               $ \\
6312.06    & [S III]  & 2.39 & $               $ & $               $ & $              $ & $  4.8\pm 1.8  $ & $               $ \\
6363.81    & [O I]    & 2.36 & $               $ & $   3.7\pm 1.0  $ & $              $ & $  3.3\pm 1.7  $ & $               $ \\
6548.06    & [N II]   & 2.26 & $               $ & $  26.7\pm 7.2  $ & $ 10.5\pm 2.9  $ & $  3.7\pm 6.2  $ & $               $ \\
6562.817   & H I      & 2.26 & $ 428.7\pm 7.9  $ & $ 596\pm 15     $ & $462.5\pm 8.4  $ & $403\pm 15     $ & $               $ \\
6583.39    & [N II]   & 2.25 & $   5.1\pm 3.4  $ & $  35.5\pm 7.4  $ & $ 37.9\pm 3.2  $ & $ 15.0\pm 6.4  $ & $               $ \\
6678.149   & He I     & 2.20 & $               $ & $   3.87\pm 0.80$ & $  4.2\pm 2.2  $ & $  3.2\pm 2.4  $ & $               $ \\
6716.42    & [S II]   & 2.17 & $               $ & $  38.1\pm 1.2  $ & $ 53.1\pm 3.1  $ & $ 26.6\pm 3.3  $ & $               $ \\
6730.78    & [S II]   & 2.16 & $               $ & $  30.7\pm 1.1  $ & $ 33.7\pm 2.8  $ & $ 26.3\pm 3.3  $ & $               $ \\
7065.179   & He I     & 2.01 & $               $ & $  13.2\pm 1.0  $ & $  5.1\pm 2.0  $ & $              $ & $               $ \\
7135.8     & [Ar III] & 1.98 & $               $ & $  16.2\pm 1.1  $ & $ 14.7\pm 2.7  $ & $ 15.7\pm 3.9  $ & $               $ \\
7319.92    & [O II]   & 1.90 & $               $ & $  12.0\pm 1.3  $ & $ <5           $ & $ <5           $ & $               $ \\
7330.19    & [O II]   & 1.90 & $               $ & $  11.0\pm 1.2  $ & $ <5           $ & $ <5           $ & $               $ \\
$E(B-V)_{\mathrm H\alpha}$ & (mag) && $ 0.38\pm 0.02$ & $ 0.67\pm 0.03$ & $ 0.44\pm 0.02$ & $ 0.32\pm 0.06$ & $              $ \\
$E(B-V)_{\mathrm H\gamma}$ & (mag) && $ 0.31\pm 0.10$ & $ 0.74\pm 0.11$ & $ 0.11\pm 0.17$ & $ 0.87\pm 0.37$ & $  0.28\pm 0.10$ \\
\enddata

\tablenotetext{a}{The scale is such that $F(\mathrm H\beta) = 100$.  If no uncertainty is quoted, the value is a $2\sigma$ upper limit to the line intensity.}
\tablenotetext{b}{pn-020, S33, pn-010, pn-012, S16, pn-017, and pn-019 are planetary nebulae while S10, KD29, Ho10, and Ho06 are \ion{H}{2} regions.}
\tablenotetext{c}{$A_1(\lambda)$ is the \citet{fitzpatrick1999} reddening law for a reddening $E(B-V) = 1$\,mag and parametrized with a ratio of total-to-selective extinction of 3.041.}

\end{deluxetable}

\begin{deluxetable}{lcllllll}
\tabletypesize{\footnotesize}
%\rotate
\tablecaption{Chemical abundances for PNe$^{\mathrm a}$ and \ion{H}{2} regions$^{\mathrm a}$ in NGC 6822\label{table_chem_abun}}
\tablewidth{0pt}
%\tablehead{
%\colhead{wavelength} & \colhead{ion} & \colhead{pn-020} & \colhead{S33} & \colhead{pn-010} &
%\colhead{pn-012} & \colhead{S16} & \colhead{pn-017} &
%}
\startdata
\hline\hline\\[1pt]
quantity     & line(s) used & pn-020                          &  S33                             & pn-010                           & pn-012                            & S16                            & pn-017                        \\[3pt]
\hline\\[1pt]
$T_e$           & 4363/5007    & $ < 15790                     $ & $   19479^{+835}_{-1253}       $ & $   19754^{+5220}_{-4183}      $ & $   14930^{+5456}_{-4877}       $ & $   14019^{+1552}_{-1441}    $ & $   21711^{+8775}_{-5866}    $\\
$N_e$           & 6716/6731    & $    2000                     $ & $    2183^{13563}_{-1784}      $ & $     186                      $ & $     685^{+84}_{-88}           $ & $    2000                    $ & $    2000                    $\\
$\mathrm{He}^0/\mathrm H$     &       5876   & $   0.080^{+0.010}_{-0.011}   $ & $   0.053^{+0.018}_{-0.014}    $ & $   0.087^{+0.026}_{-0.029}    $ & $   0.075^{+0.029}_{-0.029}     $ & $   0.083^{+0.010}_{-0.009}  $ & $   0.024^{+0.005}_{-0.006}  $\\
$\mathrm{He}^+/\mathrm H$     &       4686   & $ < 0.009                     $ & $   0.055^{+0.002}_{-0.002}    $ & $   0.034^{+0.007}_{-0.007}    $ & $   < 0.012                     $ & $   0.020^{+0.003}_{-0.002}  $ & $   0.039^{+0.013}_{-0.010}  $\\
$10^6\,\mathrm N^+/\mathrm H$ &       6584   & $0.217E^{+0.028}_{-0.025}     $ & $20.1^{+6.0}_{-1.8}            $ & $0.97^{+0.71}_{-0.40}          $ & $0.5^{+1.3}_{-0.4}              $ & $1.28^{0.69}_{-0.49}         $ & $1.6^{+1.4}_{-0.7}           $\\
$10^5\,\mathrm O^+/\mathrm H$ &       3727   & $< 1.6                        $ & $0.25^{+0.66}_{-0.13}          $ & $1.9^{+2.6}_{-1.0}             $ & $1.6^{+9.0}_{-1.2}              $ & $0.67^{+0.33}_{-2.0}         $ & $                            $\\
$10^5\,\mathrm O^+/\mathrm H$ &       7325   & $0.62^{+0.53}_{-0.66}         $ & $                              $ & $                              $ & $< 3.8                          $ & $1.3^{+1.9}_{-1.3}           $ & $< 1                         $\\
$10^5\,\mathrm O^{2+}/\mathrm H $ &   5007   & $8.0^{+1.1}_{-1.0}            $ & $6.4^{+1.2}_{-0.6}             $ & $4.2^{+3.3}_{-1.6}             $ & $8.0^{+18}_{-4.1}               $ & $14.7^{+5.5}_{-3.7}          $ & $3.3^{+3.6}_{-1.6}           $\\
$10^6\,\mathrm{Ne}^{2+}/\mathrm H $ &  3869   & $26^{+12}_{-9}                $ & $6.5^{+1.8}_{-1.1}             $ & $10^{+14}_{-6}                 $ & $23^{+105}_{-17}                $ & $30^{+20}_{-12}              $ & $6.8^{+8.9}_{-3.5}           $\\
ICF(N)       &              & $   14.88                     $ & $   41.97                      $ & $    3.95                      $ & $    6.61                       $ & $   26.53                    $ & $    8.31                    $\\
ICF(O)       &              & $    1.07                     $ & $    1.61                      $ & $    1.25                      $ & $    1.10                       $ & $    1.16                    $ & $    1.91                    $\\
ICF(Ne)      &              & $    1.15                     $ & $    1.67                      $ & $    1.82                      $ & $    1.32                       $ & $    1.21                    $ & $    2.49                    $\\
$\mathrm{He}/\mathrm H$       &              & $   0.088^{+0.010}_{-0.011}   $ & $   0.108^{+0.018}_{-0.014}    $ & $   0.121^{+0.027}_{-0.030}    $ & $   0.087^{+0.029}_{-0.029}     $ & $   0.103^{+0.010}_{-0.010}  $ & $   0.063^{+0.014}_{-0.012}  $\\
%            &              & $3.22E-06   4.13E-07  3.74E-07$ & $8.46E-04   2.51E-04   7.64E-05$ & $3.82E-06   2.81E-06   1.56E-06$ & $3.26E-06    8.66E-06   2.49E-06$ & $3.39E-05  1.83E-05  1.29E-05$ & $1.34E-05  1.13E-05  5.80E-06$\\
$12+\log(\mathrm N/\mathrm H)$ &             & $    6.51^{+0.06}_{-0.05}     $ & $    8.93^{+0.13}_{-0.04}      $ & $    6.58^{+0.32}_{-0.18}      $ & $    6.5^{+1.2}_{-0.3}          $ & $    7.53^{+0.23}_{-0.16}    $ & $    7.13^{+0.37}_{-0.19}    $\\
%            &              & $9.17E-05   1.30E-05  1.29E-05$ & $1.06E-04   2.16E-05   9.40E-06$ & $7.60E-05   5.18E-05   2.39E-05$ & $1.03E-04    2.26E-04   4.74E-05$ & $1.78E-04  6.37E-05  4.25E-05$ & $8.20E-05  6.97E-05  3.01E-05$\\
$12+\log(\mathrm O/\mathrm H)$ &             & $    7.96^{+0.06}_{-0.06}     $ & $    8.03^{+0.09}_{-0.04}      $ & $    7.88^{+0.30}_{-0.14}      $ & $    8.01^{+0.96}_{-0.20}       $ & $    8.25^{+0.16}_{-0.10}    $ & $    7.91^{+0.37}_{-0.16}    $\\
%            &              & $3.04E-05   1.36E-05  1.10E-05$ & $1.08E-05   3.05E-06   1.88E-06$ & $1.81E-05   2.60E-05   1.04E-05$ & $2.99E-05    1.39E-04   2.19E-05$ & $3.65E-05  2.38E-05  1.40E-05$ & $1.69E-05  2.22E-05  8.67E-06$\\
$12+\log(\mathrm{Ne}/\mathrm H)$ &            & $    7.48^{+0.19}_{-0.16}     $ & $    7.03^{+0.12}_{-0.08}      $ & $    7.26^{+0.62}_{-0.25}      $ & $    7.5^{+2.0}_{-0.3}          $ & $    7.56^{+0.28}_{-0.17}    $ & $    7.23^{+0.57}_{-0.22}    $\\
%            &              & $    0.04       0.03      0.04$ & $    7.95      20.93       3.97$ & $    0.05       0.08       0.03$ & $    0.03        0.20       0.03$ & $    0.19      0.14      0.09$ & $    0.16      0.14      0.07$\\
$\log(\mathrm N/\mathrm O)$   &              & $   -1.45^{+0.38}_{-0.47}     $ & $    0.90^{+1.14}_{-0.22}      $ & $   -1.30^{+0.67}_{-0.29}      $ & $   -1.5^{+2.8}_{-0.5}          $ & $   -0.72^{+0.31}_{-0.21}    $ & $   -0.79^{+0.37}_{-0.19}    $\\
%            &              & $    0.33       0.15      0.13$ & $    0.10       0.03       0.02$ & $    0.24       0.39       0.17$ & $    0.29        1.52       0.26$ & $    0.21      0.15      0.09$ & $    0.21      0.35      0.14$\\
$\log(\mathrm{Ne}/\mathrm O)$  &              & $   -0.48^{+0.20}_{-0.17}     $ & $   -0.99^{+0.15}_{-0.09}      $ & $   -0.62^{+0.71}_{-0.30}      $ & $   -0.5^{+2.3}_{-0.4}          $ & $   -0.69^{+0.33}_{-0.20}    $ & $   -0.69^{+0.75}_{-0.30}    $\\[1pt]
\hline\\[1pt]
quantity     & line(s) used & pn-019                          &  S10                             & KD29                             & Ho10                            & Ho06                            \\[3pt]
\hline\\[1pt]
$T_e$           & 4363/5007    & $   15735^{+1499}_{-1419}     $ & $   14774^{+2279}_{-2108}      $ & $   < 13872                    $ & $   12935^{+4179}_{-4465}       $ & $   10596^{+1549}_{-1638}    $\\
$N_e$           & 6716/6731    & $     100                     $ & $     188^{+143}_{-108}        $ & $     100                      $ & $     629^{+1162}_{-511}        $ & $     100                    $\\
$\mathrm{He}^0/\mathrm H$     &       5876   & $   0.174^{+0.011}_{-0.011}   $ & $   0.048^{+0.004}_{-0.005}    $ & $   0.075^{+0.009}_{-0.009}    $ & $   0.067^{+0.020}_{-0.017}     $ & $   0.078^{+0.002}_{-0.005}  $\\
$\mathrm{He}^+/\mathrm H$     &       4686   & $ < 0.002                     $ & $                              $ & $                              $ & $                               $ & $                            $\\
$10^6\,\mathrm N^+/\mathrm H$ &       6584   & $0.23^{+0.22}_{-0.16}         $ & $1.30^{+0.80}_{-0.49}          $ & $2.06^{+0.14}_{-0.14}          $ & $1.1^{+3.3}_{-0.7}              $ & $                            $\\
$10^5\,\mathrm O^+/\mathrm H$ &       3727   & $< 6                          $ & $                              $ & $5.85^{+0.34}_{-0.34}          $ & $                               $ & $6.1^{+9.2}_{-3.1}           $\\
$10^5\,\mathrm O^+/\mathrm H$ &       7325   & $                             $ & $2.1^{+4.0}_{-2.2}             $ & $< 1.6                         $ & $< 2                            $ & $                            $\\
$10^5\,\mathrm O^{2+}/\mathrm H$ &    5007   & $4.3^{+1.3}_{-0.9}            $ & $3.0^{+1.7}_{-0.9}             $ & $1.34^{+0.03}_{-0.03}          $ & $6.5^{+21}_{-3.4}               $ & $12.3^{+9.9}_{-4.3}          $\\
$10^6\,\mathrm{Ne}^{2+}/\mathrm H $ &  3869   & $6.4^{+4.1}_{-2.6}            $ & $4.0^{+4.4}_{-2.1}             $ & $                              $ & $                               $ & $                            $\\
ICF(N)       &              & $   69.53                     $ & $    2.48                      $ & $    1.23                      $ & $    4.25                       $ & $    3.00                    $\\
ICF(O)       &              & $       1                     $ & $       1                      $ & $       1                      $ & $       1                       $ & $       1                    $\\
ICF(Ne)      &              & $    1.01                     $ & $    1.68                      $ & $    5.38                      $ & $    1.31                       $ & $    1.50                    $\\
$\mathrm{He}/\mathrm H$       &              & $   0.176^{+0.011}_{-0.011}   $ & $   0.049^{+0.004}_{-0.005}    $ & $   0.075^{+0.009}_{-0.009}    $ & $   0.067^{+0.020}_{-0.017}     $ & $   0.078^{+0.002}_{-0.005}  $\\
%            &              & $1.60E-05   1.54E-05  1.14E-05$ & $3.23E-06   1.97E-06   1.20E-06$ & $2.53E-06   1.66E-07   1.72E-07$ & $4.63E-06    1.39E-05   3.00E-06$ & $                            $\\
$12+\log(\mathrm N/\mathrm H)$ &             & $    7.20^{+0.42}_{-0.31}     $ & $    6.51^{+0.26}_{-0.16}      $ & $    6.40^{+0.03}_{-0.03}      $ & $    6.67^{+1.30}_{-0.28}       $ & $                            $\\
%            &              & $4.41E-05   1.27E-05  8.92E-06$ & $5.09E-05   4.36E-05   2.35E-05$ & $7.19E-05   3.45E-06   3.38E-06$ & $8.51E-05    2.12E-04   3.42E-05$ & $1.84E-04  1.35E-04  5.30E-05$\\
$12+\log(\mathrm O/\mathrm H)$ &             & $    7.64^{+0.12}_{-0.09}     $ & $    7.71^{+0.37}_{-0.20}      $ & $    7.86^{+0.02}_{-0.02}      $ & $    7.93^{+1.08}_{-0.17}       $ & $    8.27^{+0.32}_{-0.12}    $\\
%            &              & $6.46E-06   4.11E-06  2.61E-06$ & $6.65E-06   7.39E-06   3.47E-06$ & $                              $ & $                               $ & $                            $\\
$12+\log(\mathrm{Ne}/\mathrm H)$ &            & $    6.81^{+0.28}_{-0.18}     $ & $    6.82^{+0.48}_{-0.23}      $ & $                              $ & $                               $ & $                            $\\
%            &              & $    0.36       0.35      0.26$ & $    0.06       0.13       0.07$ & $    0.04       0.00       0.00$ & $    0.05        0.16       0.04$ & $                            $\\
$\log(\mathrm N/\mathrm O)$   &              & $   -0.44^{+0.42}_{-0.31}     $ & $   -1.20^{+0.89}_{-0.48}      $ & $   -1.45^{+0.04}_{-0.04}      $ & $   -1.26^{+1.30}_{-0.28}       $ & $                            $\\
%            &              & $    0.15       0.10      0.07$ & $    0.13       0.16       0.08$ & $                              $ & $                               $ & $                            $\\
$\log(\mathrm{Ne}/\mathrm O)$  &              & $   -0.83^{+0.30}_{-0.20}     $ & $   -0.88^{+0.54}_{-0.26}      $ & $                              $ & $                               $ & $                            $\\
\enddata

\tablenotetext{a}{pn-020, S33, pn-010, pn-012, S16, pn-017, and pn-019 are planetary nebulae while S10, KD29, Ho10, and Ho06 are \ion{H}{2} regions.}

\end{deluxetable}

\begin{deluxetable}{lcc}
\tabletypesize{\scriptsize}
%\rotate
\tablecaption{Atomic data used\label{table_atomic_data}}
\tablewidth{0pt}
\tablehead{
\colhead{ion} & \colhead{Transition Probabilities} & \colhead{Collision Strengths}
}
\startdata
N$^+$     & \citet{wiesseetal1996} & \citet{lennonburke1994} \\
O$^+$     & \citet{wiesseetal1996} & \citet{pradhan1976} \\
          &                        & \citet{mclaughlinbell1993} \\
O$^{2+}$  & \citet{wiesseetal1996} & \citet{lennonburke1994} \\
Ne$^{2+}$ & \citet{mendozazeippen1982} & \citet{butlerzeippen1994} \\
          & \citet{kauffmansugar1986} \\
\enddata
\end{deluxetable}

\begin{deluxetable}{lcccc}
\tabletypesize{\scriptsize}
%\rotate
\tablecaption{Oxygen Abundances in \ion{H}{2} regions and PNe in Dwarf Irregulars\label{table_di_prop}}
\tablewidth{0pt}
\tablehead{
\colhead{Galaxy} & 
\colhead{$12+\log(\mathrm O/\mathrm H)_{\mathrm{H\ II}}$} &
\colhead{$12+\log(\mathrm O/\mathrm H)_{\mathrm{PN}}$} & 
\colhead{$\Delta\log(\mathrm O/\mathrm H)^{\mathrm a}$} &
\colhead{Data$^{\mathrm b}$}
}
\startdata
Leo A     & $7.38\pm 0.10$ & $7.34\pm 0.10$ & $-0.04\pm 0.14$ & 1 \\
Sextans A & $7.60\pm 0.20$ & $7.84\pm 0.11$ & $+0.29\pm 0.23$ & 2,3 \\
SMC       & $8.03\pm 0.10$ & $8.13\pm 0.06$ & $+0.10\pm 0.12$ & 4,5 \\
NGC 6822  & $8.10\pm 0.10$ & $8.01\pm 0.14$ & $-0.09\pm 0.17$ & 6,7,8 \\
Sextans B & $8.12\pm 0.12$ & $7.95\pm 0.16$ & $-0.17\pm 0.20$ & 2,3 \\
LMC       & $8.35\pm 0.06$ & $8.24\pm 0.10$ & $-0.11\pm 0.12$ & 4,5 \\
\enddata
\tablenotetext{a}{$\Delta\log(\mathrm O/\mathrm H) = \log(\mathrm O/\mathrm H)_{\mathrm{PN}} - \log(\mathrm O/\mathrm H)_{\mathrm{H\ II}}$}
\tablenotetext{b}{Data: (1) \citet{vanzeeetal2006}; (2) \citet{magrinietal2005}; (3) \citet{kniazevetal2005}; (4) \citet{russelldopita1992}; (5) a compilation of \citet{alleretal1987}, \citet{barlow1987}, \citet{monketal1987}, \citet{woodetal1987}, \citet{dopitaetal1988}, \citet{meatheringhametal1988}, \citet{henryetal1989}, \citet{meatheringhametal1988}, \citet{meatheringhamdopita1991a} , \citet{meatheringhamdopita1991b}, \citet{dopitameatheringham1991}, \citet{vassiliadisetal1992}, \citet{jacobykaler1993}, \citet{leisydennefeld1996}: see \citet{richer1993}, \citet{richermccall1995}, and \citet{stasinskaetal1998} for more details; (6) \citet{peimbertetal2005}; (7) \citet{leeetal2006}; (8) this study (planetary nebulae only)}
\end{deluxetable}

\end{document}